\documentclass[12pt,twoside,english]{article}
\usepackage{bookman}
\usepackage[T1]{fontenc}
\usepackage[latin1]{inputenc}
\usepackage{amsmath}
\usepackage{color}
\usepackage{graphicx}
\usepackage{amssymb}

\makeatletter

\providecommand{\LyX}{L\kern-.1667em\lower.25em\hbox{Y}\kern-.125emX\@}

\usepackage{bookman}
\usepackage{a4}
\usepackage{cite}
\input{epsfig.sty}
\def\fnum@table{\tablename~{\bf\thetable}}
\def\fnum@figure{\figurename~{\bf\thefigure}}
\def\tablename{\footnotesize{\bf Table}}
\def\figurename{\footnotesize{\bf Figure}}

\AtBeginDocument{

}

\usepackage{babel}
\makeatother
\begin{document}
\vspace{-0.5cm}
\begin{center}\textbf{\huge Proton-proton and deuteron-gold collisions
at RHIC}\end{center}{\huge \par}

\begin{center}{\Large Klaus Werner}$^{(a)}$%
\footnote{Invited talk, given at the XXth Winter Workshop on Nuclear Dynamics,
Trelawny Beach, Jamaica, March 2004.%
}{\Large , Fuming Liu$^{(b,c)}$ , Tanguy Pierog$^{(d)}$}\end{center}{\Large \par}

{\footnotesize $^{(a)}$ SUBATECH, Université de Nantes -- IN2P3/CNRS
-- Ecole des Mines, Nantes, France }\\
{\footnotesize $^{(b)}$ Institute of Particle Physics , Huazhong
Normal University, Wuhan, China}\\
{\footnotesize $^{(c)}$ Institut für Theoretische Physik, J. W.
Goethe Universitäat, Robert-Mayer-Str. 10, 60054 Frankfurt am Main,
Germany}\\
{\footnotesize $^{(d)}$ FZK, Institut für Kernphysik, Karlsruhe,
Germany}{\footnotesize \par}

\begin{abstract}
We try to understand recent data on proton-proton and deuteron-gold
collisions at RHIC, employing a modified parton model approach.
\end{abstract}

\section{Interesting data from RHIC}

\begin{figure}[htb]
\begin{center}\includegraphics[  scale=0.3,
  angle=270,
  origin=c]{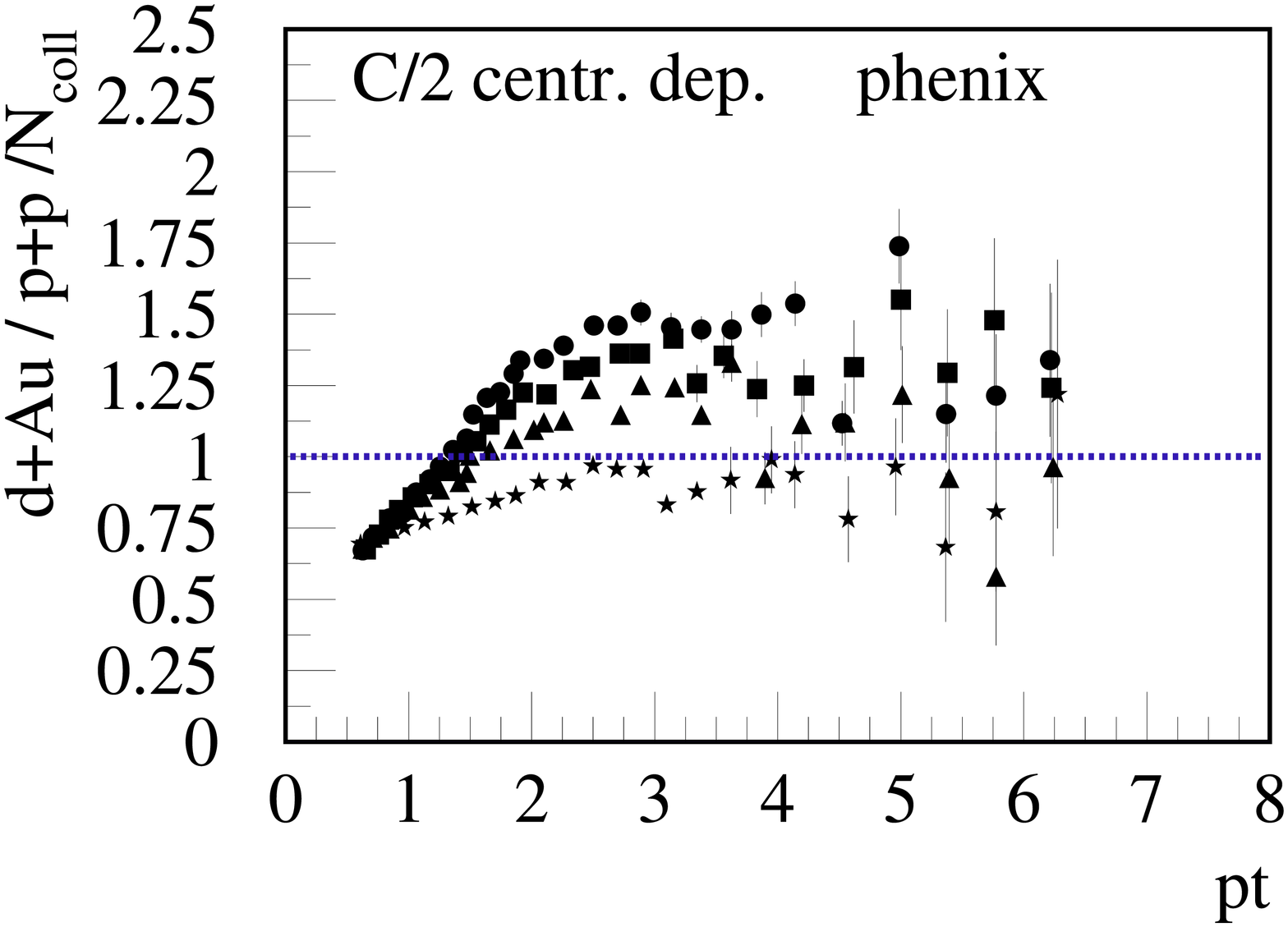}$\quad $\includegraphics[  scale=0.3,
  angle=270,
  origin=c]{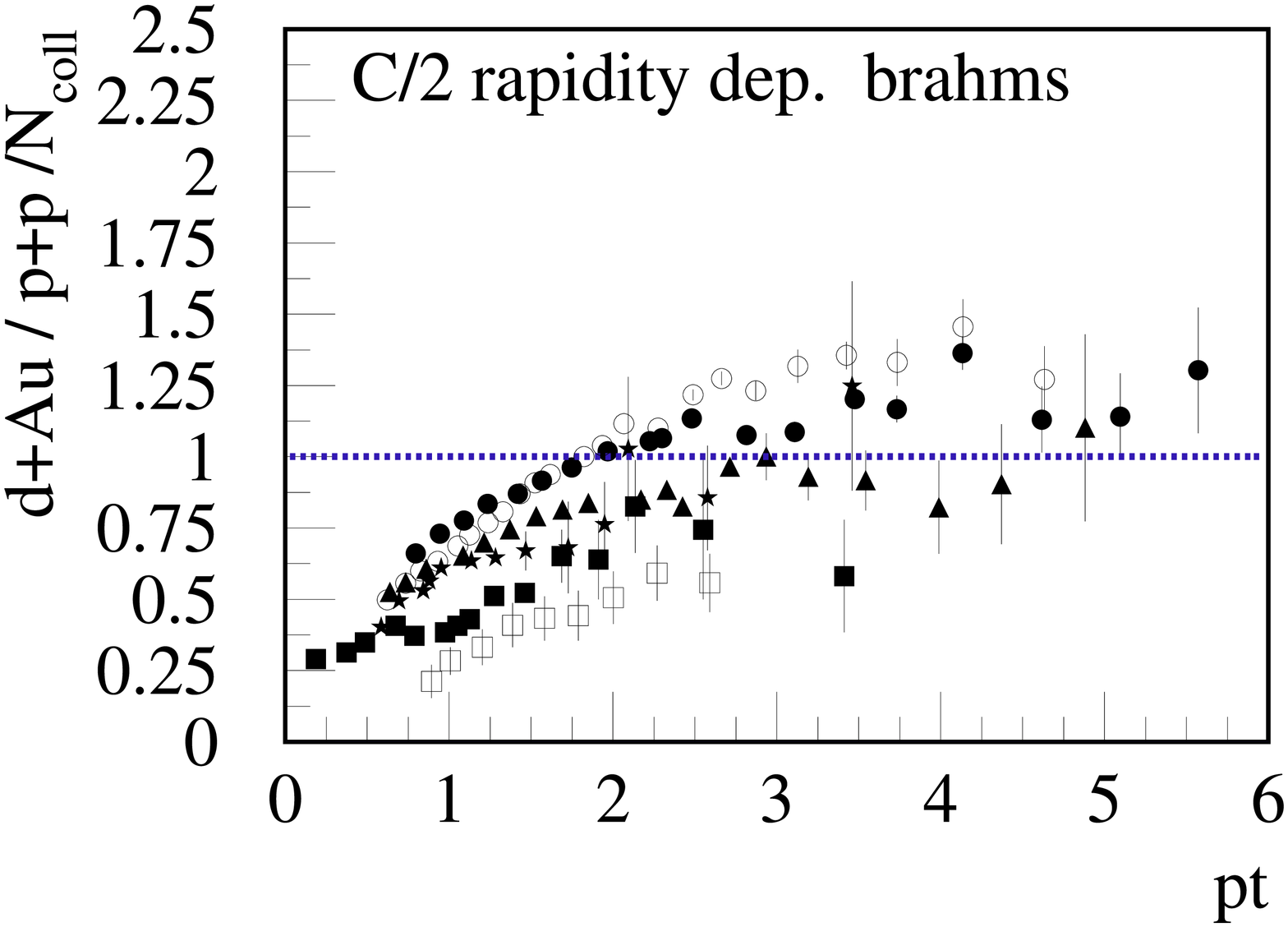}\end{center}
\vspace{-2cm}

\caption{Left: Centrality dependence of the nuclear modification factor. Top
to bottom: 0-20\%, 20-40\%, 40-60\%, 60-88\%. Right: Rapidity dependence
of the nuclear modification factor. Top to bottom: eta = 0, 1, 2.2,
3.2. \label{cap:1}}
\end{figure}

The nuclear modification factor $R=(dN_{AA}/d^{2}p_{t})\, /\, (N_{\mathrm{coll}}\, dN_{pp}/d^{2}p_{t})$
shows interesting features: For AuAu, it is much smaller than one
for central collisions, whereas for d-Au, it is bigger than one for
central collisions. With decreasing centrality, the modification factor
for d-Au approaches one and goes even below one. Also interesting
is the fact that R decreases with increasing pseudo-rapidity (see
fig \ref{cap:1}, data from \cite{phenix-dau-r-c,brahms-dau-r-rap}).

\section{High parton densities}

In this chapter we want to discuss theoretical implications of high
parton densities.

\begin{figure}[htb]
\begin{center}\begin{minipage}[c]{0.10\textwidth}%
\includegraphics[  scale=0.7]{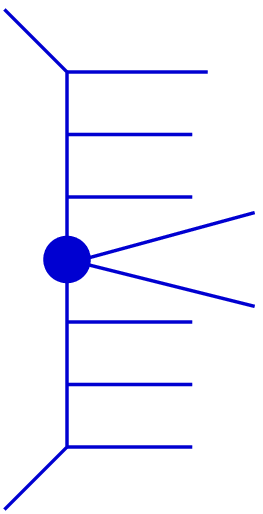}\end{minipage}%
\begin{minipage}[c]{0.10\textwidth}%
$\quad \to \quad $\end{minipage}%
\begin{minipage}[c]{0.10\textwidth}%
\includegraphics[  scale=0.7]{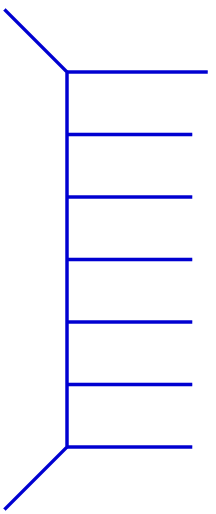}\end{minipage}%
\end{center}

\caption{Parton-parton scattering. \label{cap:Parton-parton}}
\end{figure}

Let us first consider parton-parton scattering. A parton from the
projectile, after emitting several (initial state) partons, interacts
with a corresponding parton from the target, see figure \ref{cap:Parton-parton}(left).
To simplify further discussions, we use a symbolic parton ladder for
this diagram, as shown in figure \ref{cap:Parton-parton}(right).

\begin{figure}[htb]
\begin{center}\includegraphics[  scale=0.7]{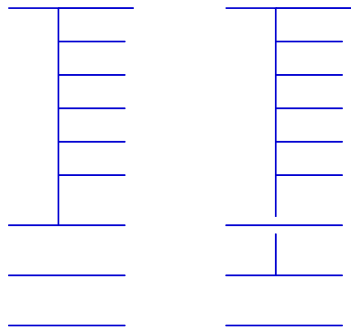}\end{center}

\caption{Scattering with many partons. \label{cap:manypartons}}
\end{figure}

Having several partons available, the projectile parton may interact
in this way with any of the target partons, as shown in fig. \ref{cap:manypartons},
and vice versa, this will simply change the cross section by some
factor.

\begin{figure}[htb]
\begin{center}(A)\includegraphics[  scale=0.7]{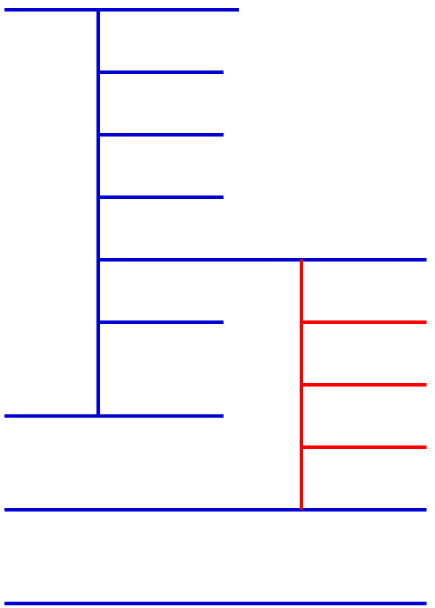}$\qquad $(B)\includegraphics[  scale=0.7]{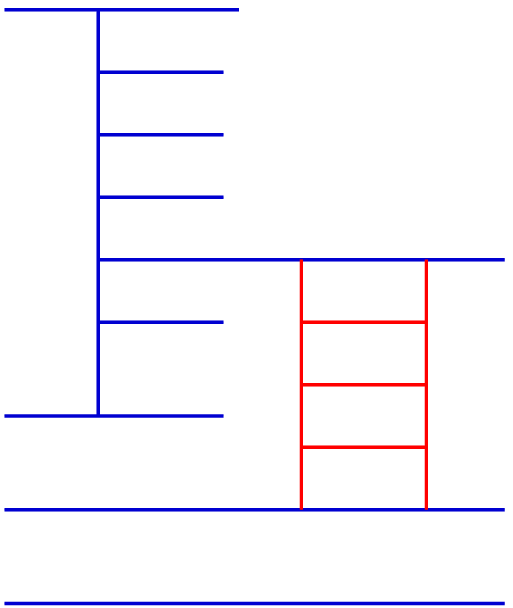}$\qquad $(C)\includegraphics[  scale=0.7]{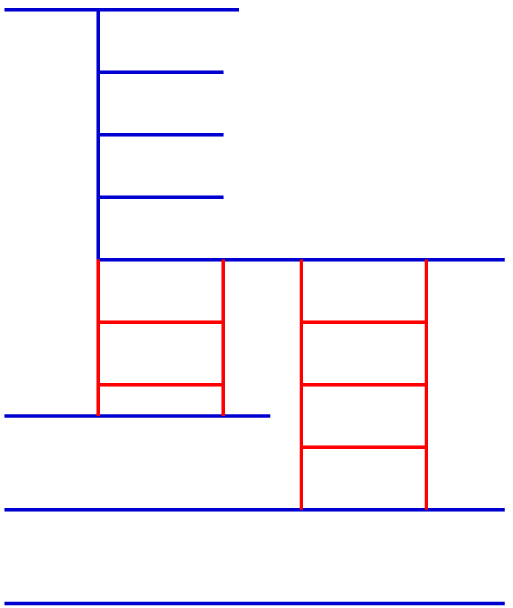}\end{center}

\caption{Multiple ladders (left), elastic interacting (middle), rapidity gap
events (right).\label{cap:hiden}}
\end{figure}

The situation will, however, be more complicated in case of high parton
densities. Here, a parton from a ladder may rescatter with another
target from the projectile or target, providing an additional ladder
(fig. \ref{cap:hiden}, left). A ladder parton may also interact elastically
(fig. \ref{cap:hiden}, middle). And finally, a parton ladder may
be linked to two closed ladders (fig. \ref{cap:hiden}, right), providing
a rapidity gap on the projectile or target side.%
Diagram (B) will interfere with the simple diagram (fig. \ref{cap:manypartons}),
and gives a negative contribution to the cross section, providing
screening. Diagram (C) can be referred to as high mass diffraction.
Particularly interesting is diagram (A), the case of multiple parton
ladders, as we are going to discuss later.\\

So we try to put all this together

\begin{figure}[htb]
\begin{center}\begin{minipage}[c]{0.35\textwidth}%
\includegraphics[  scale=0.7]{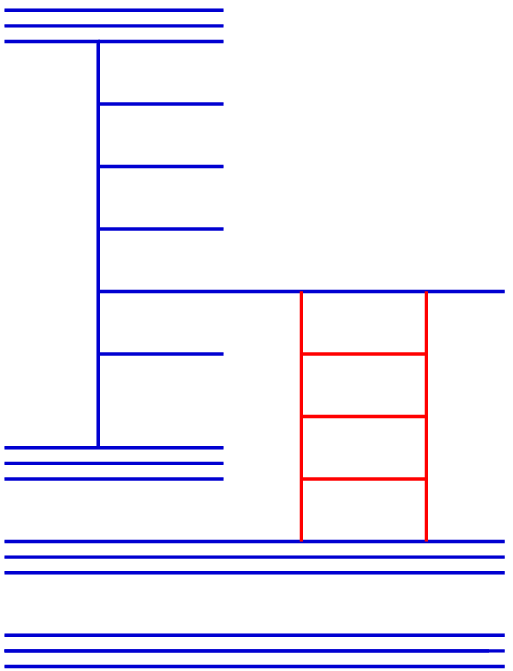}\end{minipage}%
$\quad \to \quad $\begin{minipage}[c]{0.15\textwidth}%
\includegraphics[  scale=0.7]{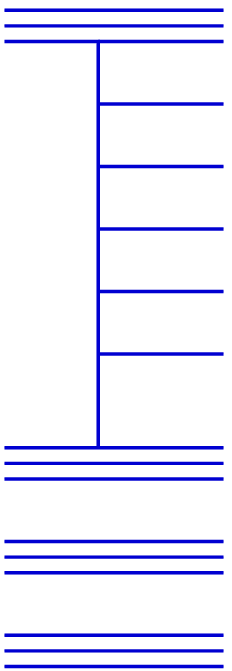}\end{minipage}%
$\qquad $\begin{minipage}[c]{0.25\textwidth}%
\textcolor{blue}{With }\\
\textcolor{blue}{reduced }\\
\textcolor{blue}{weight}\end{minipage}%
\end{center}

\caption{Screening contribution: treated as simple parton scattering, but
with reduced weight, to be referred to as screening correction. \label{cap:eff1}}
\end{figure}

\begin{figure}[htb]
\begin{center}\begin{minipage}[c]{0.35\textwidth}%
\includegraphics[  scale=0.7]{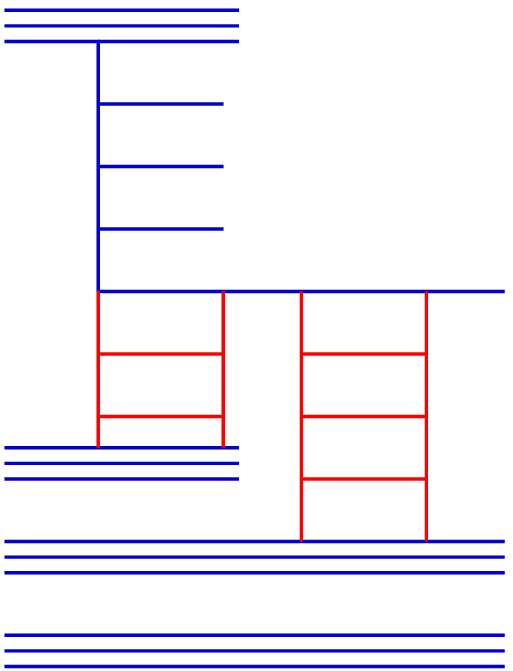}\end{minipage}%
$\quad \to \quad $\begin{minipage}[c]{0.35\textwidth}%
\includegraphics[  scale=0.7]{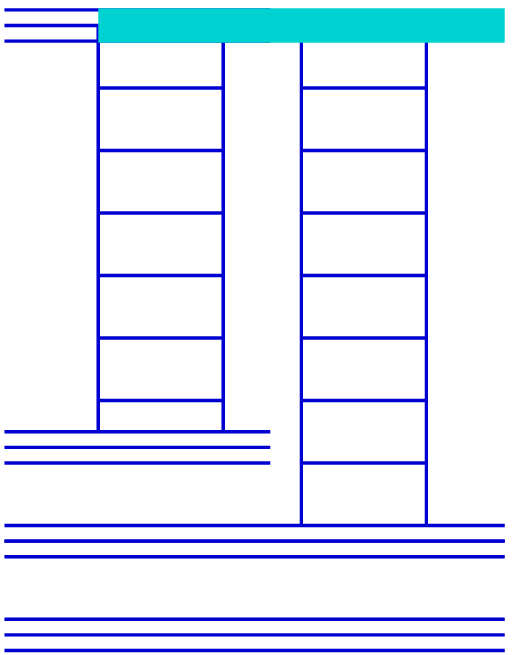}\end{minipage}%
\end{center}

\caption{The diffractive contribution: The fat line represents a projectile
excitation.\label{cap:eff2}}
\end{figure}

\begin{figure}[htb]
\begin{center}\begin{minipage}[c]{0.35\textwidth}%
\includegraphics[  scale=0.7]{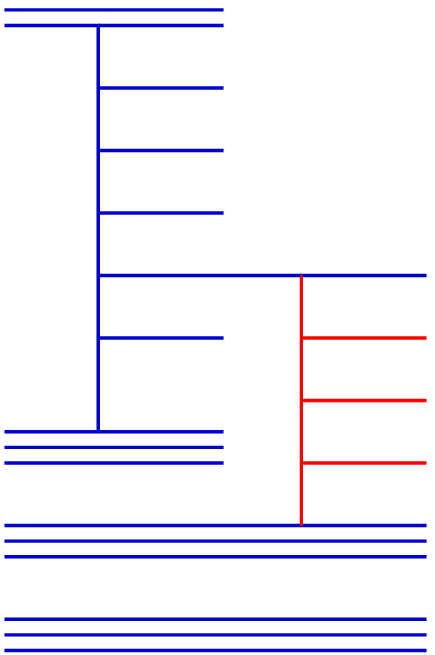}\end{minipage}%
$\quad \to \quad $\begin{minipage}[c]{0.15\textwidth}%
\includegraphics[  scale=0.7]{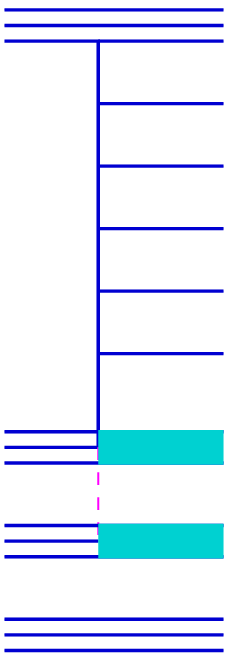}\end{minipage}%
\end{center}

\caption{Multiple ladder contribution: A target excitation may represent several
ladders.\label{cap:eff3}}
\end{figure}

\begin{itemize}
\item In a simple and transparent way
\item using just simple ladders between projectile and target (Pomerons)
\item putting all complications into {}``projectile / target excitations'',
to be treated in an effective way
\end{itemize}
as shown in figs. \ref{cap:eff1}, \ref{cap:eff2}, \ref{cap:eff3}.
Bifurcation of parton ladders will not be treated explicitly, they
are absorbed into target and projectile excitations, visualized as
fat lines in the figures. The excitations may represent one, two,
or even more ladders, depending on the parton densities.

\textbf{~}

\vspace{1cm}
\textbf{How to realize the screening correction?}

A simple diagram is roughly given as $(x^{+})^{\beta }(x^{-})^{\beta '}$,
and adding the screening diagram will reduce the contribution, which
we realize via $(x^{+})^{\beta }(x^{-})^{\beta '+\varepsilon }$ with
some positive parameter $\varepsilon $, which suppresses small $x$.
$\varepsilon $ should increase with the number $Z$ of {}``close''
partons, so $\varepsilon $ should be a monotonically increasing function
of $Z$. What are close partons? Their number $Z$ should increase
with decreasing $b$ and it should increase with energy. We use \[
Z_{P/T}=\sum _{\mathrm{nucleons}}\frac{E}{E_{0}}\, g(\frac{b}{b_{0}})\]
 with\[
g(x)=\frac{1}{\sqrt{a^{2}+x^{2}}}\exp (-x^{2}),\]
and\[
\varepsilon =\varepsilon _{\max }\big (1-\frac{1}{\sqrt{1-\big (\log (1+3\, Z)\big )^{2}}}\big )\]
\\

\textbf{How to realize projectile / target excitations ?} (accounting
for multiple, interacting (?) ladders) 

\begin{itemize}
\item We suppose an excitation mass distributed according to $1/M^{2\alpha }$.
\item For masses exceeding hadron masses we take strings.\\

\item String properties are supposed to depend on $Z$ \\
\\
(the string represents multiple, interacting (?) ladders)
\end{itemize}
For the moment we take $\left\langle p_{t}\right\rangle _{\mathrm{break}}=\left\langle p_{t}\right\rangle _{\mathrm{break}}^{0}\, f(Z)$
with $f(Z)=\min (f_{\max },1+\alpha \, Z)$, $f_{\max }=3,\: \alpha =0.3$,
which gives $Z\lesssim 2$ for pp, $Z\lesssim 6$ for d-Au.

The formalism is based on cut diagram techniques, strict energy conservation,
and Markov chains for the numerics \cite{nexus}.

\section{Some results}

\begin{figure}[htb]
\includegraphics[  scale=0.3,
  angle=270,
  origin=c]{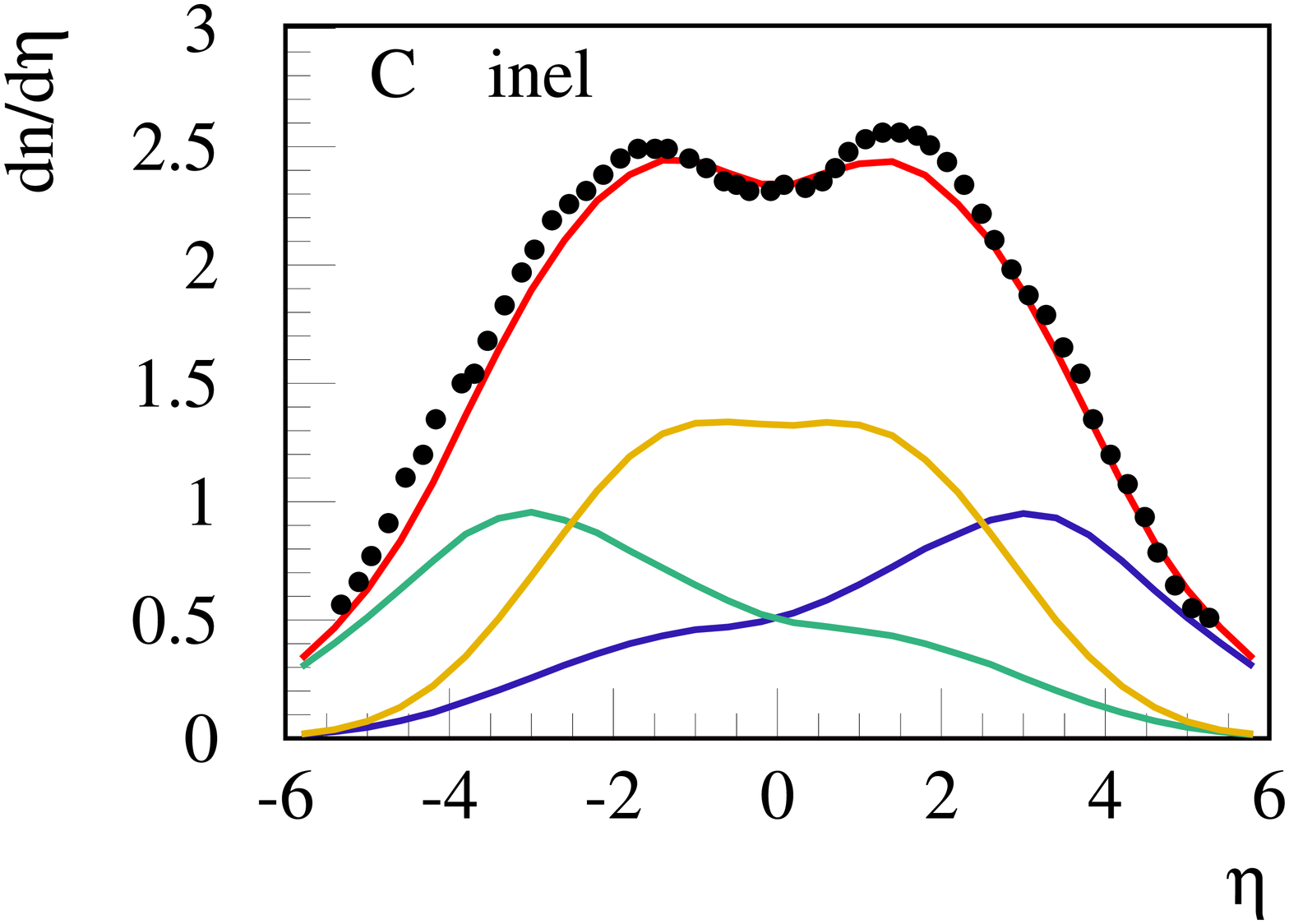}$\qquad $\includegraphics[  scale=0.3,
  angle=270,
  origin=c]{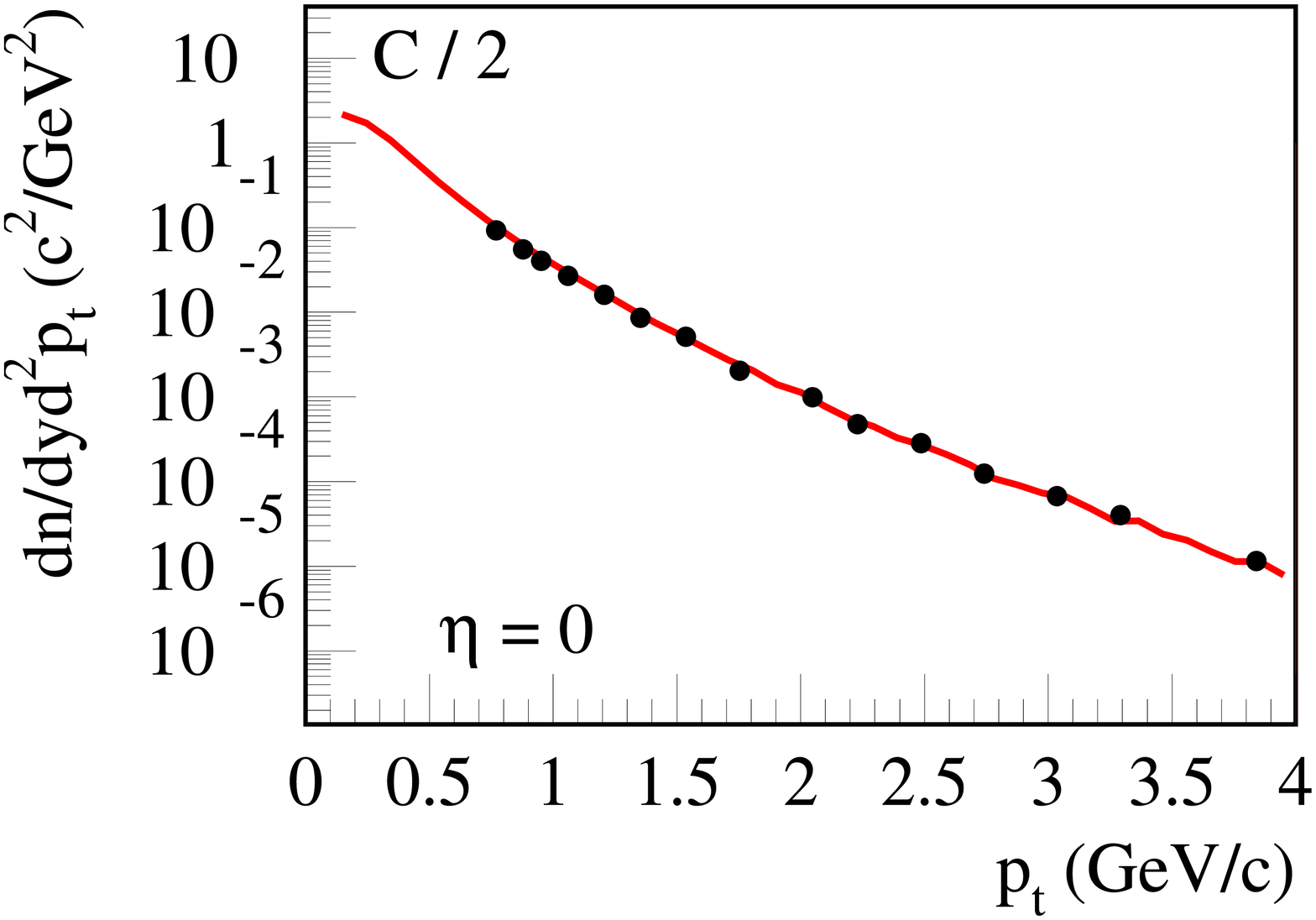}

\caption{Left: The different contributions (central ladders, target excitations,
projectile excitations) to the pseudo-rapidity distribution. Right:
Transverse momentum distribution.\label{cap:prap}}
\end{figure}

In fig.\ref{cap:prap} we show the different contributions to the
pseudo-rapidity distribution in pp collisions, compared to data from
\cite{phobos}, and a transverse momentum distribution of charged
particles (data from \cite{brahms-dau-r-rap}). %

\begin{figure}[htb]
\includegraphics[  scale=0.3,
  angle=270,
  origin=c]{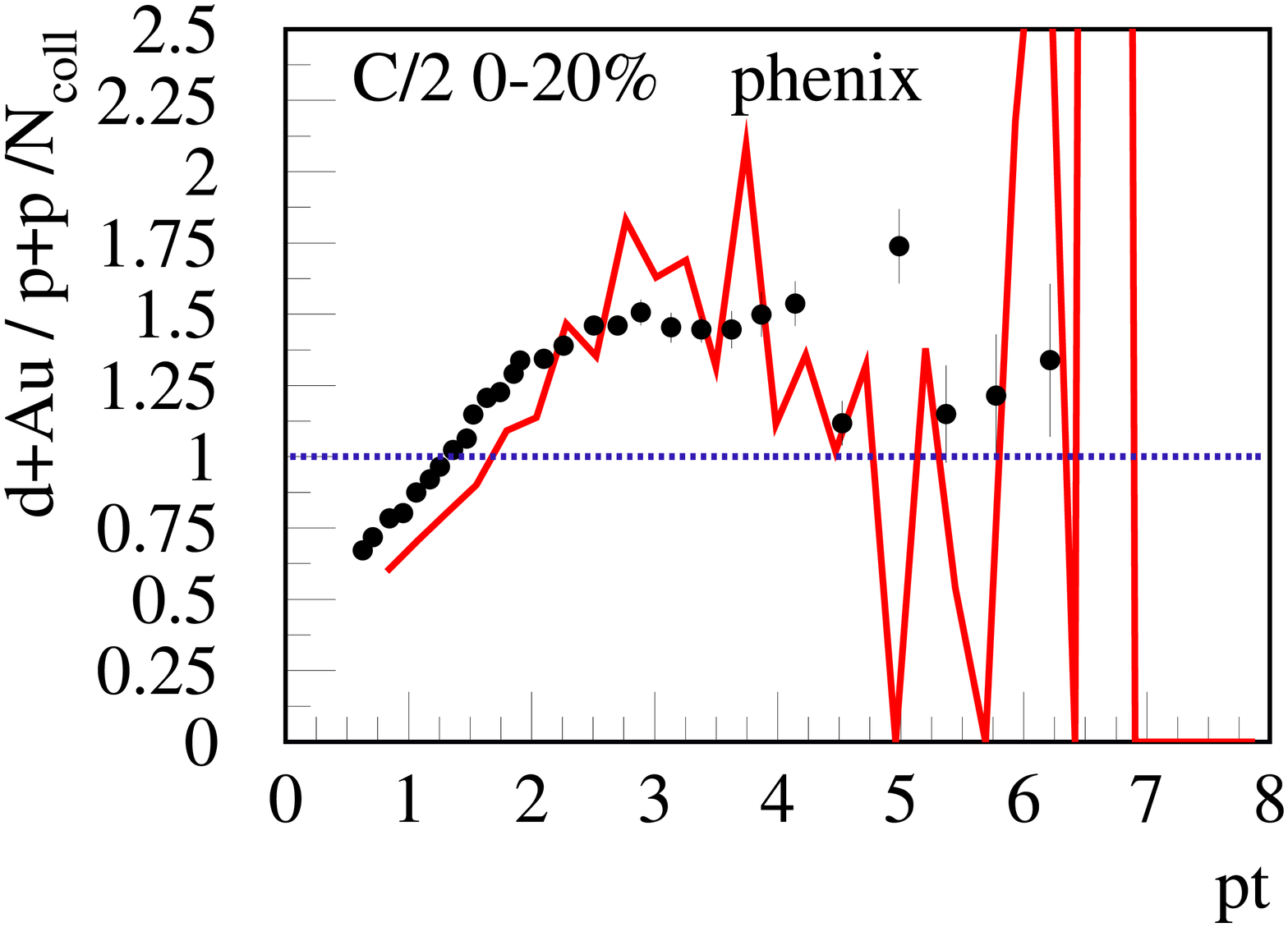}\includegraphics[  scale=0.3,
  angle=270,
  origin=c]{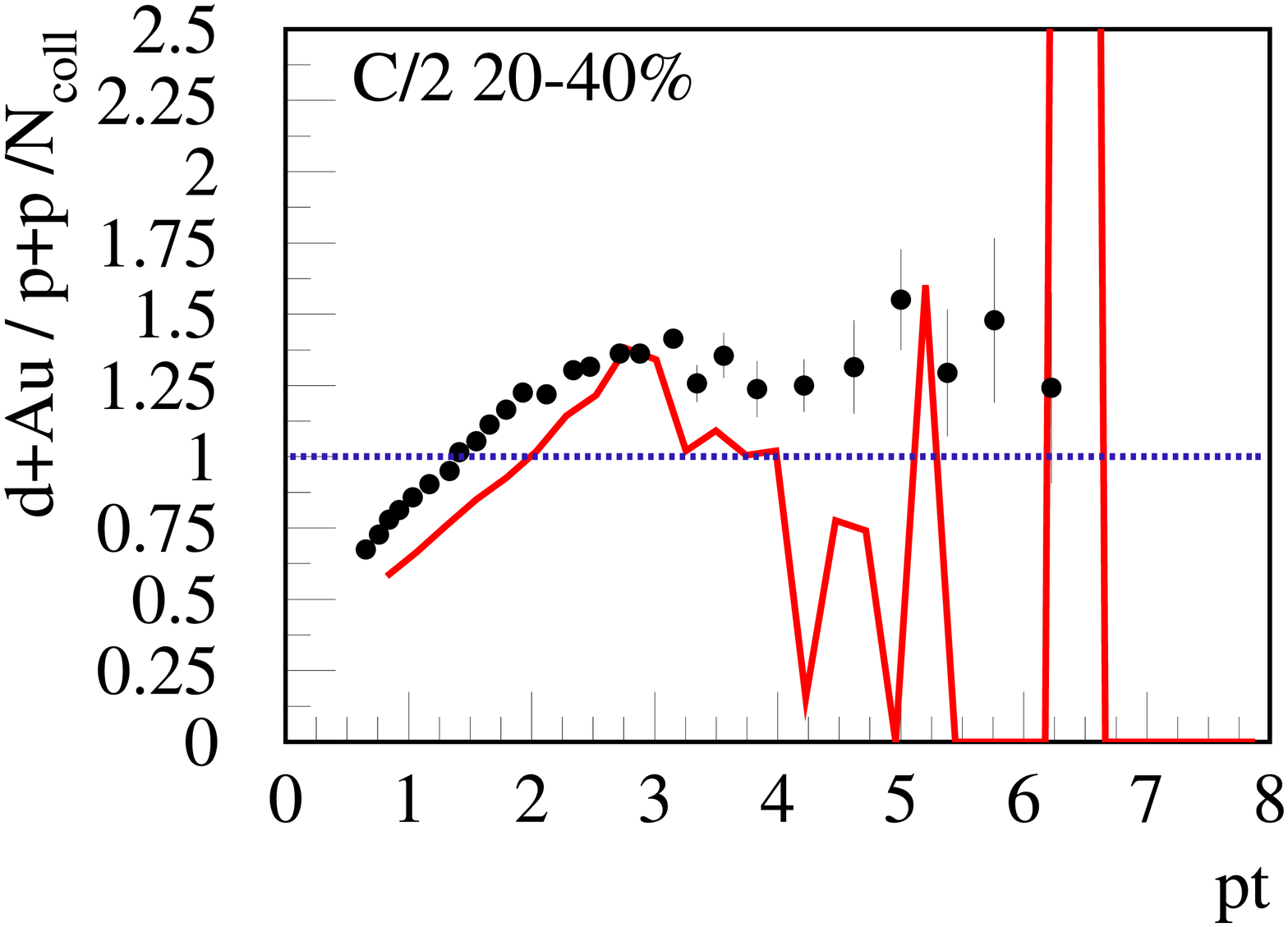}

\includegraphics[  scale=0.3,
  angle=270,
  origin=c]{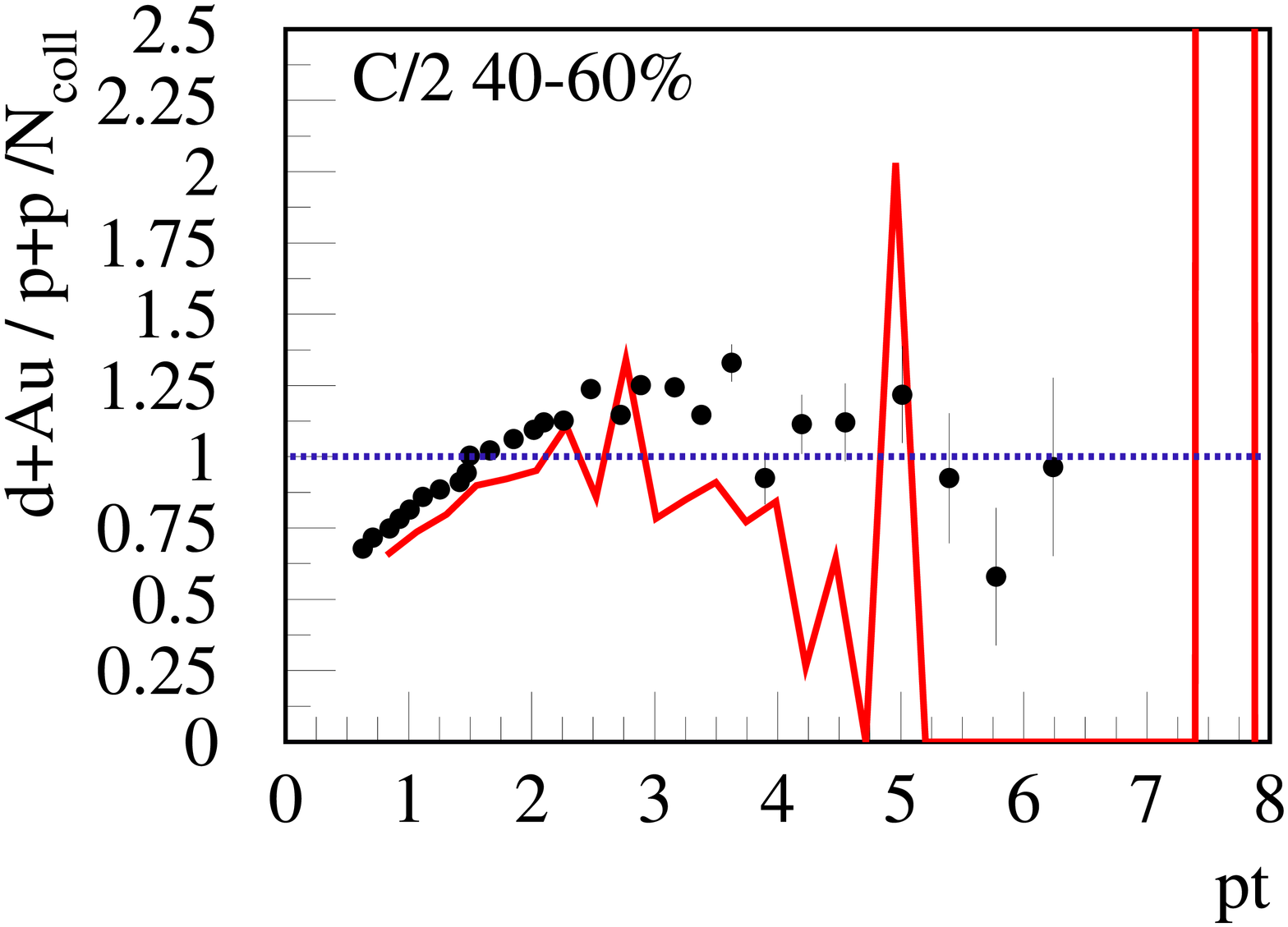}\includegraphics[  scale=0.3,
  angle=270,
  origin=c]{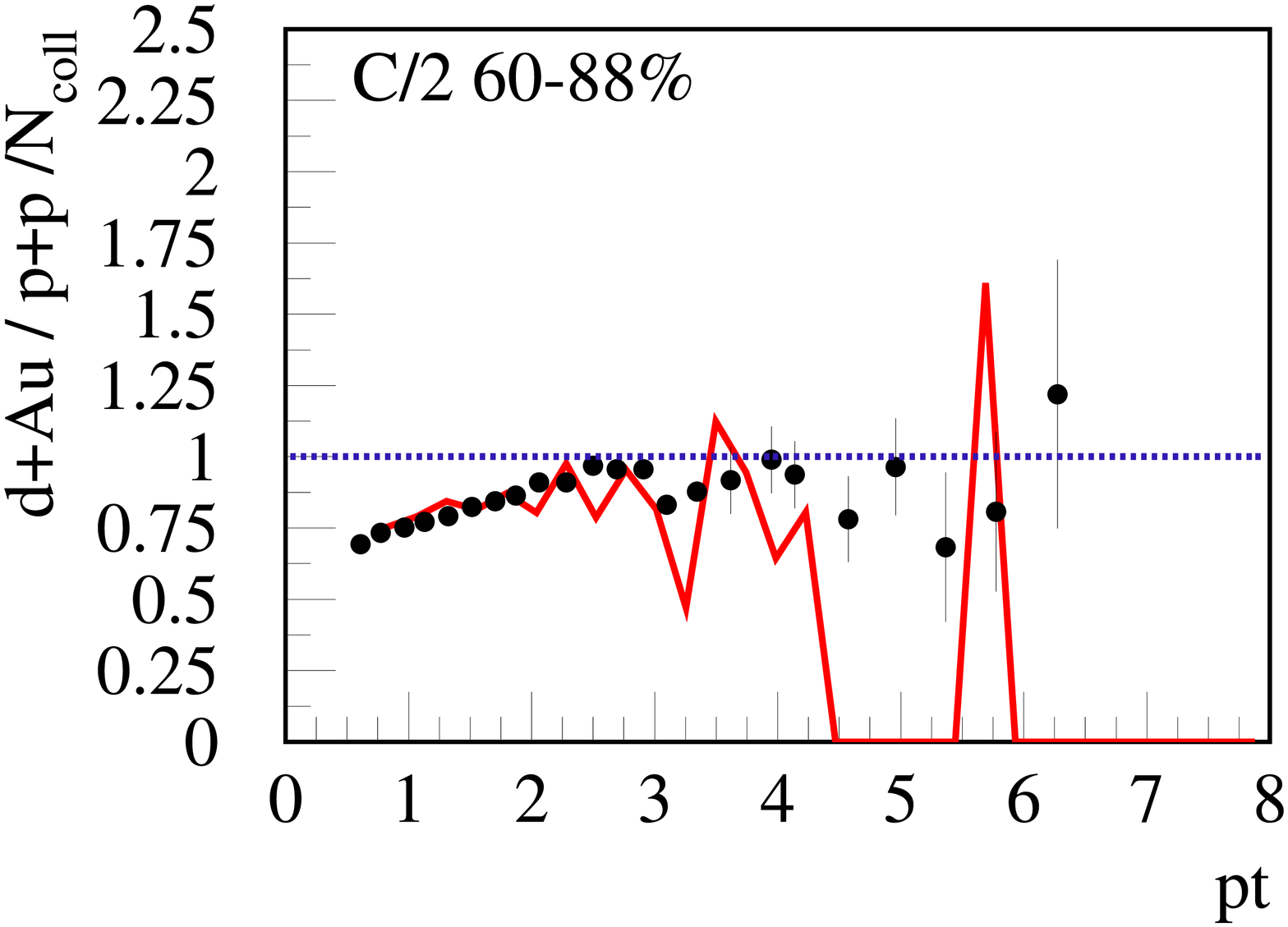}

\caption{Centrality dependence of the nuclear modification factor. \label{cap:cent}}
\end{figure}

\begin{center}%
\begin{figure}[htb]
\includegraphics[  scale=0.3,
  angle=270,
  origin=c]{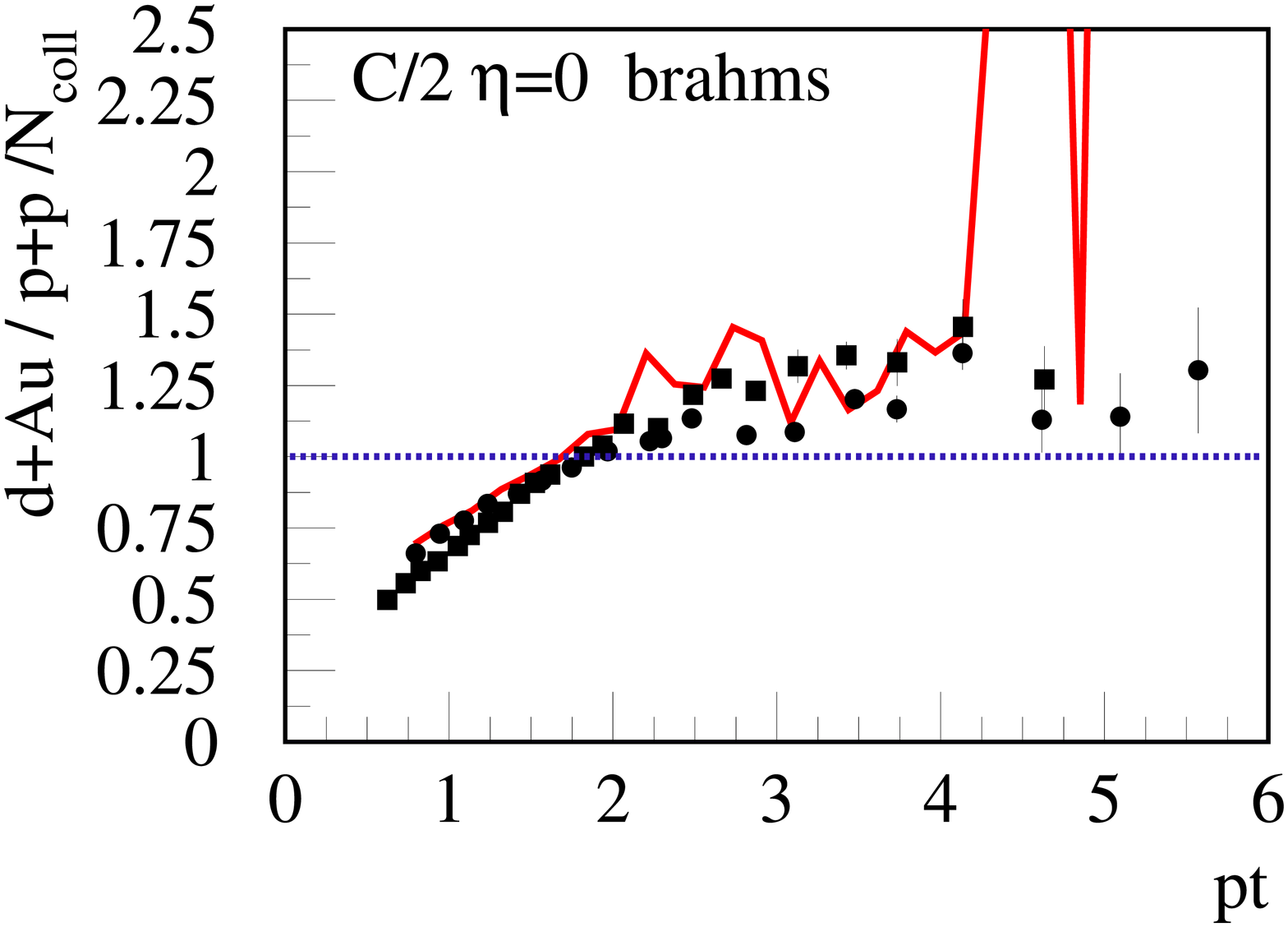}\includegraphics[  scale=0.3,
  angle=270,
  origin=c]{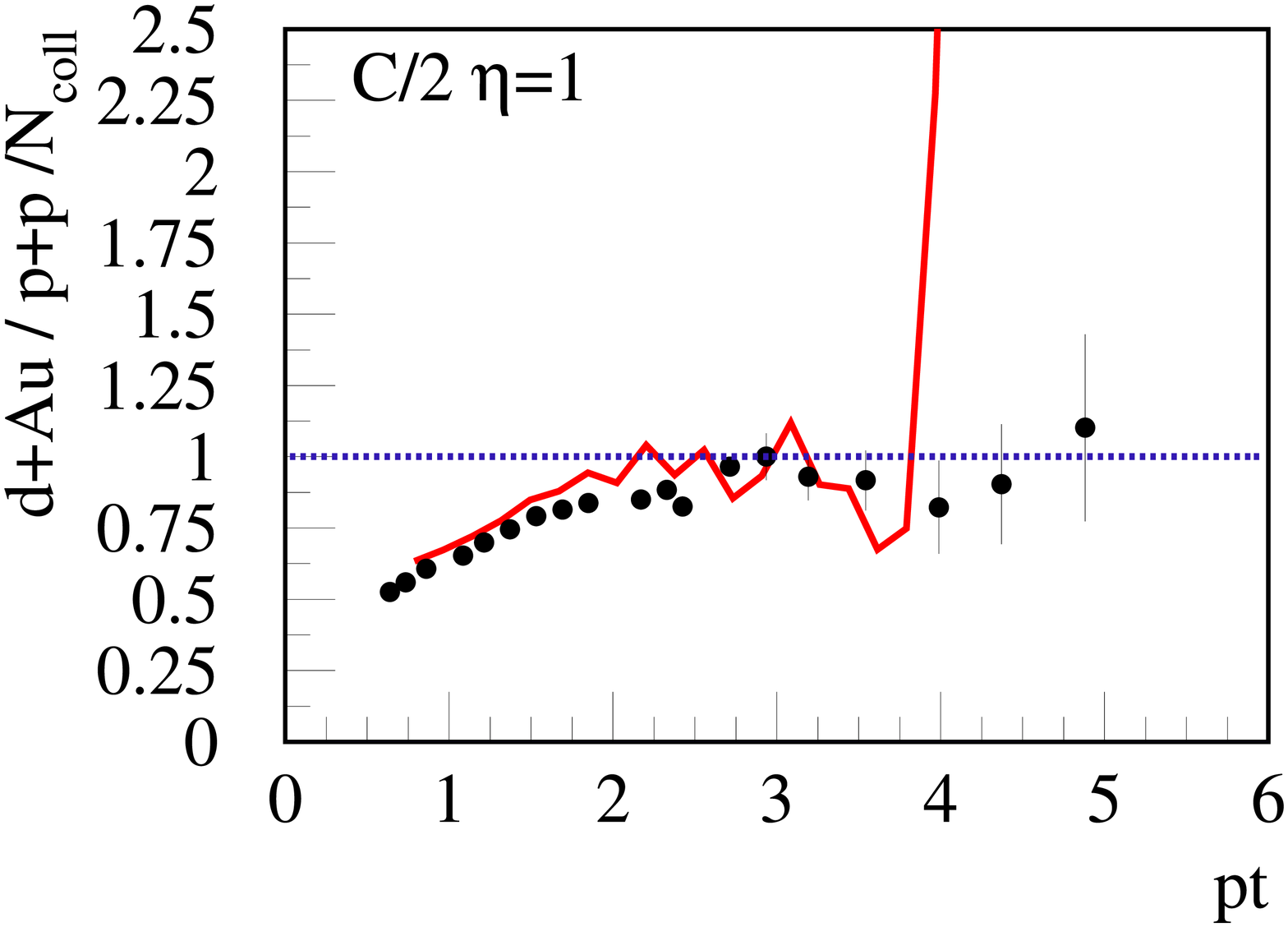}

\includegraphics[  scale=0.3,
  angle=270,
  origin=c]{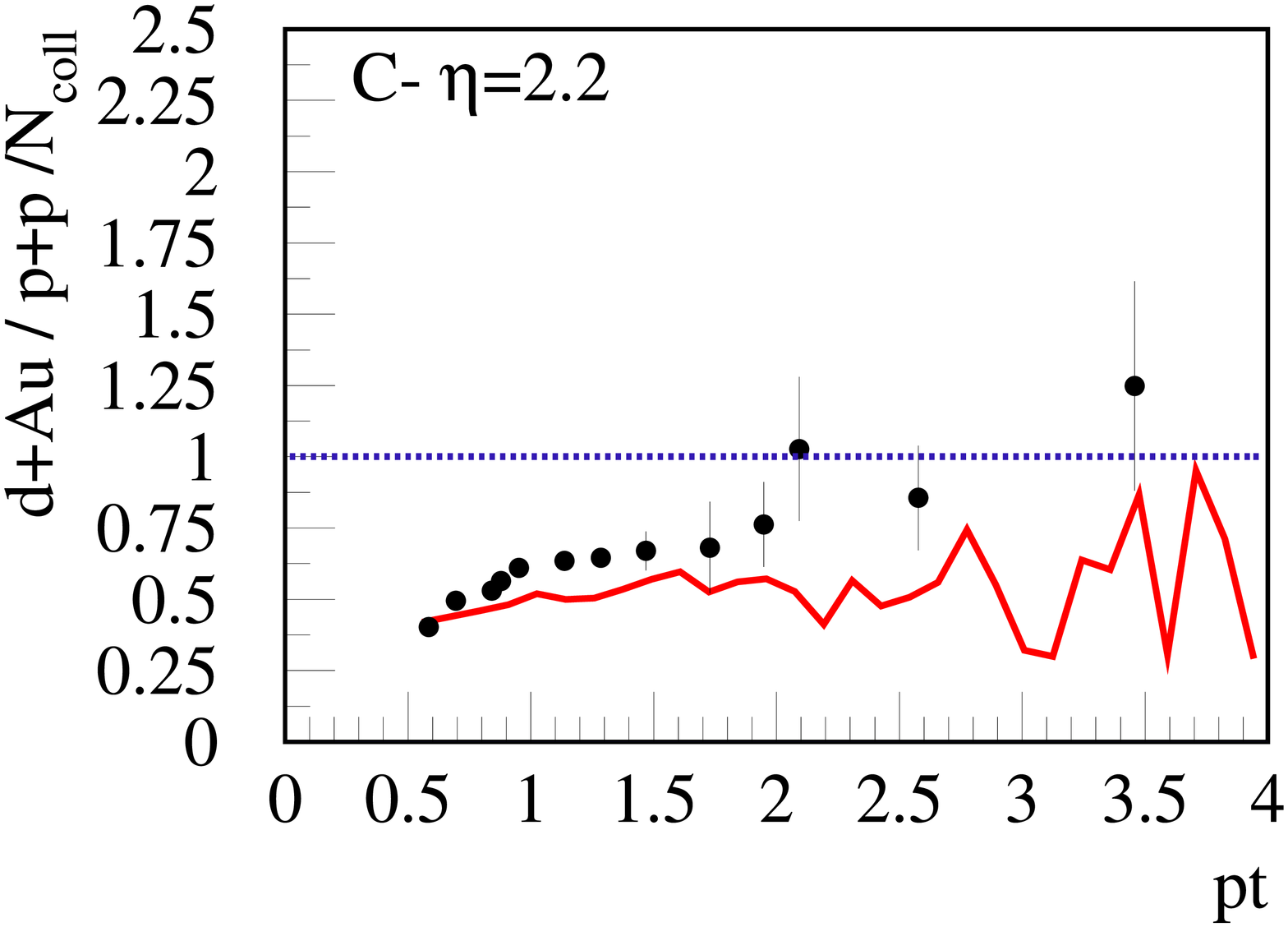}\includegraphics[  scale=0.3,
  angle=270,
  origin=c]{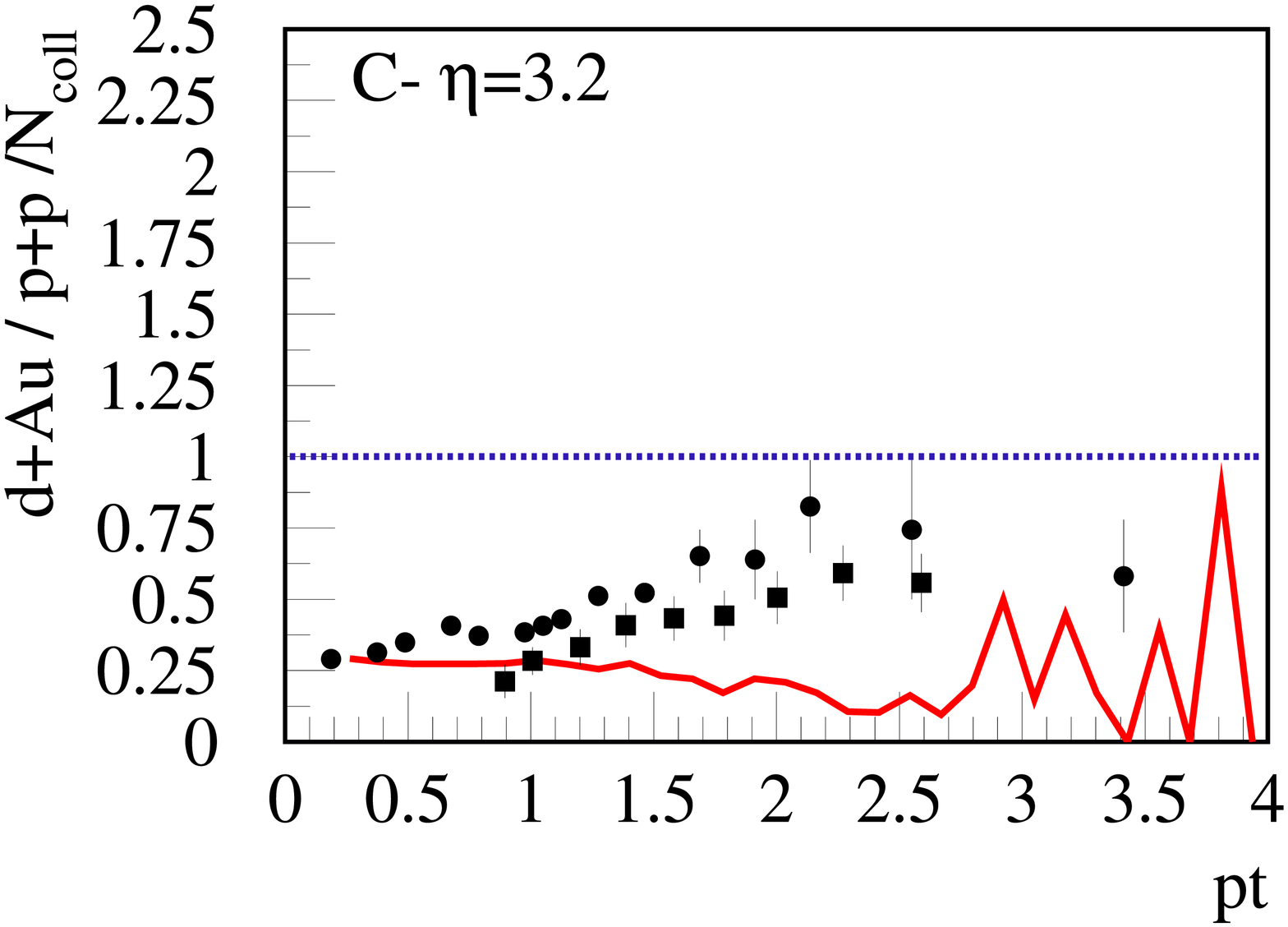}

\caption{Pseudo-rapidity dependence of the nuclear modification factor. \label{cap:rap}}
\end{figure}
\end{center}

Let us turn to d-Au scattering. At this point there is no fine-tuning
employed. We first want to understand the qualitative features of
our effective treatment of interacting parton ladders  (projectile
/ target excitations). In figs. \ref{cap:cent}, \ref{cap:rap} we
show the centrality and the pseudo-rapidity dependence of the nuclear
modification factor, showing clearly the effect of an increased transverse
momentum due to interacting ladders.

\end{document}